\documentclass[useAMS,usenatbib,referee]{mn2e}{{
\usepackage{subeqnarray}
\usepackage{graphicx}
\usepackage{natbib}
\include{epsf} 
\include{rotate}

\title{Molecular Hydrogen Formation on Porous Dust Grains}

\author[H. B. Perets, O. Biham] 
{Hagai B. Perets$^{1}$ and Ofer Biham$^{2}$  \\ 
$^{1}$Center for Astrophysics, Weizmann Institute of Science, 
Rehovot 76100, Israel \\ 
$^{2}$Racah Institute of Physics, The Hebrew University, Jerusalem 91904, Israel 
} 

\begin{document}

\maketitle

\begin{abstract}

Recent laboratory experiments on interstellar dust analogues
have shown that H$_2$ formation on dust grain surfaces is efficient
in a range of grain temperatures below 20 K.
These results indicate that surface
processes may account for the observed H$_2$ abundance in cold diffuse
and dense clouds.
However, high abundances of H$_2$ have also been observed in warmer
clouds, including photon-dominated regions (PDRs), 
where grain temperatures may reach 50 K,
making the surface processes extremely inefficient.
It was suggested that this apparent discrepancy can be resolved
by chemisorption sites.
However, recent experiments indicate that chemisorption processes
may not be efficient at PDR temperatures. 
Here we consider the effect of grain porosity on H$_2$ formation, and analyze
it using a rate equation model. 
It is found that porosity extends the efficiency of
the recombination process to
higher temperatures. 
This is because H atoms that desorb from the internal surfaces of
the pores may re-adsorb many times and thus stay longer on the surface.
However, this porosity-driven extension  
may enable efficient H$_2$ formation
in PDRs only if porosity also contributes to significant cooling
of the grains, compared to non-porous grains.

\end{abstract}

\begin{keywords}
ISM: molecules - molecular processes
\end{keywords}

\section{Introduction} 
\label{sec:intro}

H$_2$ is the most abundant molecule in the Universe.
It plays a key role in interstellar chemistry and serves as a coolant
during gravitational collapse and star formation.
Therefore, the processes of formation and dissociation of molecular hydrogen
strongly affect the evolution of interstellar clouds
\citep{Duley1984,Williams1998}. 
Already in the 1960's it was found that H$_2$ cannot form efficiently 
enough in the gas phase to account for its observed abundances,
and it was suggested that dust grains serve as catalysts for this process 
\citep{Gould1963}. 
The formation of molecular hydrogen on dust grain surfaces was studied,
assuming that H atoms that collide with a grain are  
physically adsorbed (physisorbed).
These atoms diffuse on the surface and form H$_2$ molecules
when they encounter each other
\citep{Hollenbach1970,Hollenbach1971,Hollenbach1971b}.
The formation rate 
$R_{\rm H_2}$ (cm$^{-3}$s$^{-1}$), 
of H$_2$ molecules can be expressed by

\begin{equation}
R_{\rm H_2} = 
n_{\rm H} v_{\rm H} \sigma n_{\rm grain} \xi \eta / 2
\end{equation}

\noindent
where 
$n_{\rm H}$  (cm$^{-3}$) 
is the number density of hydrogen atoms in the gas,
$v_{\rm H} = \sqrt{ 8k_B T_g/(\pi m_H) }$
(cm s$^{-1}$)
is their typical velocity
(where 
$m_H=1.67 \cdot 10^{-24}$ gram
is the mass of an H atom
and $T_g$ 
is the gas temperature),
$n_{\rm grain}$ (cm$^{-3}$) 
is the number density of dust grains
and
$\sigma$ (cm$^2$)
is the average grain cross section
\citep{Hollenbach1971b}. 
Assuming spherical grains, 
$\sigma = \pi r^2 $ 
where $r$ (cm) is the grain radius.
The parameter  
$0 \le \xi \le 1$ 
is the sticking coefficient
and the parameter
$0 \le \eta \le 1$ 
is the recombination efficiency,
namely, the fraction of adsorbed atoms which 
come out in molecular form.

The total grain mass
amounts to about $1\%$ of the 
hydrogen mass in the cloud.
The number density of dust grains is 
$n_{\rm grain} \simeq 10^{-12} n$
where
$n=n_{\rm H} + 2 n_{\rm H_2}$ 
(cm$^{-3}$) 
is the total density of hydrogen atoms
in both atomic and molecular ($n_{\rm H_2}$) forms
\citep{Hollenbach1971b}. 
Assuming compact grains with mass density of 
$\rho_g = 2$ (gram cm$^{-3}$),
these parameters lead to a typical grain diameter of 
about $0.2 \mu$m. 
Assuming gas temperature of 100 K
as well as 
$\xi \simeq 1$
and
$\eta \simeq 1$,
one obtains

\begin{equation}
R_{\rm H_2} \simeq R  n_{\rm H} n,
\end{equation}

\noindent
where
the rate constant is
$R \simeq 10^{-17}$ (cm$^{3} s^{-1})$,
in agreement with observations for diffuse clouds
\citep{Jura1975}.
More recent observations in photon dominated regions,
indicate that in these regions the rate coefficient
should be
$R \simeq 10^{-16}$ (cm$^3$ s$^{-1}$),
in order to account for the observed 
H$_2$ abundance
\citep{Habart2003}.

Recent laboratory experiments indicate that
a model based on physisorbed atoms may account for the H$_2$ formation
rate in cold interstellar clouds, where the grain temperature does
not exceed 20 K.
However, molecular hydrogen has also been observed
in large abundances in photon-dominated regions
(PDRs), which are exposed to high fluxes of untra-violet radiation
\citep{Tielens1997}. 
In these regions, typical grain temperatures are in the range of
30 - 50 K. 
In order to explain the formation of H$_2$ in PDRs
it was suggested that some of the H atoms are 
chemically adsorbed (chemisorbed), 
and thus may remain on the surface long enough to form molecules
even at higher grain temperatures. 
However, recent experimental and theoretical
results indicate that such chemisorption processes 
are likely to contribute to H$_2$ formation only at grain temperatures
above 450 K. 
Such high grain temperatures can be achived by ultra-small
grains with diameters of the order of 1 nm, 
which may exhibit large temperature
fluctuations 
\citep{Draine1985,Guhathakurta1989}.
However, the efficiency of H$_2$ production on such 
extremely small grains is expected to be low.
This is due to 
the very small population of H atoms on each grain,
where the recombination rate is dominated by the discreteness
and fluctuations in the population of adsorbed H atoms
\citep{Lipshtat2005}. 
Thus, chemisorption processes may be 
irrelevant to the range of grain temperatures observed in PDRs. 

In this paper we consider the effect of porosity on the rate
of H$_2$ formation on interstellar dust grains.
We present a model in which H atoms can be adsorbed either on the
external surface of the grains or inside the pores.
The model shows that H atoms
can be retained in pores
for longer times,
and recombine there to form H$_2$ molecules at higher temperatures than 
on non-porous grains.

The paper is organized as follows.
In Sec. 
\ref{sec:compact}
we consider the formation of H$_2$ molecules on compact grains,
which is efficient only within a narrow temperature range
below 20 K.
In Sec. 
\ref{sec:problem} 
we describe the high H$_2$ abundance in
PDRs, which cannot be explained by the processes considered
in Sec.
\ref{sec:compact}.
In Sec. 
\ref{sec:evidence} 
we review some of the evidence that interstellar dust grains
are porous. 
A model for H$_2$ formation on porous grains is presented
in Sec. 
\ref{sec:model}. 
It is shown that porosity extends the efficiency range of
H$_2$ formation toward higher grain temperatures. 
The results are discussed in Sec.
\ref{sec:discussion} 
and summarized 
in Sec. 
\ref{sec:summary}. 

\section{H$_2$ formation on dust grains with compact shape}
\label{sec:compact}

During the past decade,
the formation of molecular hydrogen on analogues of interstellar dust
materials has been studied by laboratory experiments.
In particular, carbonaceous samples
\citep{Pirronello1997a,Pirronello1997b,Vidali1998a,Vidali1998b,Pirronello1999,Zecho2002a,Zecho2002b,Perry2003,Guttler2004,Zecho2004}, 
silicates such as olivine
\citep{Pirronello1997a,Pirronello1997b,Vidali1998a} 
and a variety of amorphous ice samples 
\citep{Manico2001,Roser2001,Roser2002,Hornekaer2003,Hornekaer2005,Perets2005}
have been studied. 
The results of some of these experiments were analyzed
using rate equation models. 
The relevant parameters, namely
the energy barriers for the atomic 
diffusion
and 
desorption
and for molecular desorption, 
have been found.
These parameters were used in order to calculate the efficiency 
of molecular recombination 
under conditions relevant to the interstellar clouds. 
It was found that the recombination of H atoms physisorbed on
these surfaces is efficient within a narrow temperature range,
below 20 K
\citep{Katz1999,Cazaux2004,Perets2005}.

Consider an interstellar dust grain exposed to a flux of
H atoms.
The formation of molecular hydrogen on the grain surface
is described by

\begin{equation}
{dN \over dt} = F - W N - 2 A N^2,
\label{eq:rate}
\end{equation}

\noindent
where $N$ is the number of H atoms on the grain.  
The first term in 
Eq.
(\ref{eq:rate})
describes 
the effective flux, $F$ (s$^{-1}$), 
of H atoms  
sticking to
the surface, 
given by
$F = n_{\rm H} v_{\rm H} \sigma \xi$. 
The Langmuir-Hinshelwood (LH) rejection of atoms deposited on top of already
adsorbed hydrogen atoms (or molecules) is neglected here, since the coverage
is assumed to be low.
The second term describes the thermal 
desorption of H atoms from the surface
where

\begin{equation}
W = \nu \exp(-{E_1/k_{\rm B}T})
\label{eq:W}
\end{equation}

\noindent
is the desorption rate,
$\nu$ is the attempt rate 
[standardly taken as $10^{12}$ (s$^{-1}$)],
$E_1$ is the activation energy barrier for desorption and 
$T$ (K) is the grain temperature.
The third term in 
Eq.~(\ref{eq:rate}) 
accounts for the depletion of adsorbed H atoms 
due to formation of H$_2$ molecules. 
The parameter $A = a/S$ is the rate at which H atoms scan the entire
surface of the grain, where 

\begin{equation}
a = \nu \exp(-{E_0/k_{\rm B}T})
\label{eq:A}
\end{equation}

\noindent
is the hopping rate between adjacent adsorption sites and
$E_0$ is the energy barrier for hopping. 
For a spherical grain
the number of adsorption sites on the surface 
is 
$S=4\pi r^2 s$ 
where
$s$ (cm$^{-2}$)
is their density.
The formation rate, 
$R_{\rm grain}$, 
of H$_2$ on a single grain is given by
$R_{\rm grain} = A N^2$.
Here we assume, for simplicity, that all H$_2$ molecules desorb
from the surface upon formation.
Under steady state conditions one can obtain an exact solution
for $R_{\rm grain}$
in terms of $F$, $W$ and $A$
\citep{Biham1998}.
The recombination efficiency 

\begin{equation}
\eta = \frac{2 R_{\rm grain}}{F}
\end{equation}

\noindent
turns out to exhibit strong temperature dependence. 
Taking into account the Langmuir rejection
term, it is found that H$_2$ recombination is efficient
at grain temperatures in the range
$T_0 \le T \le T_1$
where
\citep{Biham2002}

\begin{equation}
T_0 = {E_0 \over {k_B \ln (\nu S / F)}}
\end{equation}

\noindent
and

\begin{equation}
T_1 = { {2 E_1 - E_0} \over {k_B \ln (\nu S / F)}}. 
\label{eq:gvul12}
\end{equation}

\noindent
The simulations presented in this paper are based
on the parameters obtained
experimentally for amorphous carbon,
namely the activation energies are 
$E_{0}=44.0$ meV
and
$E_{1}=56.7$ meV
\citep{Katz1999},
and the density of adsorption sites on the
surface is
$s \simeq 5 \times 10^{15}$
(cm$^{-2}$)
\citep{Biham2001}.
For simplicity,
the sticking coefficient is assumed to be
$\xi = 1$.
The gas phase density is
$n_{\rm H}=100$ (cm$^{-3}$)
and its temperature is
$T_g=500$ K,
thus
$v_H=3.24 \times 10^5$ (cm s$^{-1}$).
Under these conditions
one obtains that for amorphous carbon the range of grain temperatures
in which H$_2$ formation is efficient is
$12 \le T \le 16$ K.

\section{The abundance of H$_2$ in photon dominated regions}
\label{sec:problem}

The temperatures of both the gas and dust grains
in PDRs
are significantly higher than in other H I regions. 
This is due to
far ultraviolet radiation which penetrates the clouds,
heats the gas through photon adsorption
and the dust through photoelectric heating 
\citep{Hollenbach1997}.
The temperatures of interstellar grains
in several PDR environments, such as 
Chamaeleon, Oph W, S140, IC63, NGC2023 and the Orion Bar,
were recently calculated 
\citep{Habart2004}. 
An analytic expression 
for the temperatures of large grains
\citep{Hollenbach1991}
was used,
and the median temperature of small
grains was evaluated. 
It was found that in most of 
these regions the {\it average} temperatures of small grains
are in the range of 10 - 30 K (though temperature fluctuations may heat 
them to temperatures higher than 100 K for very short durations), 
while the temperatures 
of large grains are between 30 - 50 K.
Observations show that in all these PDRs, 
the rate constant of H$_2$ formation is 
$3 \cdot 10^{-17} \le R \le 1.5 \cdot 10^{-16}$
(cm$^3$ s$^{-1}$).
This is a rather high formation rate,
which is comparable with that obtained in colder, diffuse clouds. 
In order to achieve such rate, the recombination efficiency 
must be very high, namely
$\eta \simeq 1$.

The process of H$_2$ formation on compact grains described above
is not efficient at these high grain temperatures and cannot account for the
observed abundance.
This is because at these temperatures
the residence time of H atoms on the grain
is too short to enable them to encounter each other and recombine.
In order to explain the formation efficiency at 
higher temperatures it was suggested
\citep{Hollenbach1971}  
that the grain surface 
contains enhanced adsorption sites with 
``semi-chemical binding'' sites
where atoms can stick much stronger to the surface. 
As a result they retain on the
surface for a longer period, that enables more efficient 
recombination.
It was suggested that these enhanced binding sites are
chemical adsorption (chemisorption) sites.
Their binding energies
were calculated, mainly for carbonaceous dust materials 
\citep{Bennet1971,Dovesi1976,Dovesi1981,Aronowitz1985}. 
However, recent
theoretical and experimental results indicate that chemisorption sites
play an important role only at much higher surface temperatures.
Theoretical studies have shown that
H atoms on graphite need to penetrate an
energy barrier of 
$\simeq$ 0.2 eV 
in order to reach a chemisorption site on basal planes
\citep{Jeloaica1999,Sha2002a,Sha2002b}.
This result was confirmed experimentally when an
activation energy barrier of 0.18 eV was
found experimentally for hydrogen on C(0001) 
(Zecho, private communication). 
The gas temperatures prevalent in PDRs may enable H atoms from the
gas phase to enter chemisroption sites upon collision with a grain.
However, it is unlikely for atoms already adsorbed in physisorption
sites (and thus equilibrated with the surface)
to penetrate this barrier and hop into chemisorption sites.

Chemisorbed H atoms may reside in defect sites, in step edges
(for which the energy barrier for penetration 
is negiligible) or in basal plane sites. 
The recombination of these atoms  
takes place through two possible processes.
In the LH process, a desorbing H atom picks another H
atom from an adjacent site to form a molecule.
In the Eley-Rideal (ER)
mechanism, an atom from the gas phase abstarcs an adsorbed atom to
from a molecule. 
However, the ER mechanism is relevant only at high 
surface coverage, which is not likely to exist 
under interstellar conditions.
The LH process occurs only at surface temperatures higher than ~440 K for
basal plane sites, and at even higher temperatures around
~850 K for step 
edge and defect sites 
\citep{Zecho2002a,Zecho2002b,Guttler2004,Zecho2004}. 
 
The relevance of these results to interstellar dust materials
is thus not yet clear.
If interstellar dust materials exhibit these features,
it may indicate that in PDRs
(where typical grain temperatures are between 30-50 K), 
chemisorption processes do not contribute significantly
to H$_2$ formation on dust grain surfaces 
\citep{Cazaux2004,Habart2004}. 

\section{Evidence for Porosity in Interstellar Dust Grains}
\label{sec:evidence}

\begin{center}
\begin{figure} 
\includegraphics[width=6in]{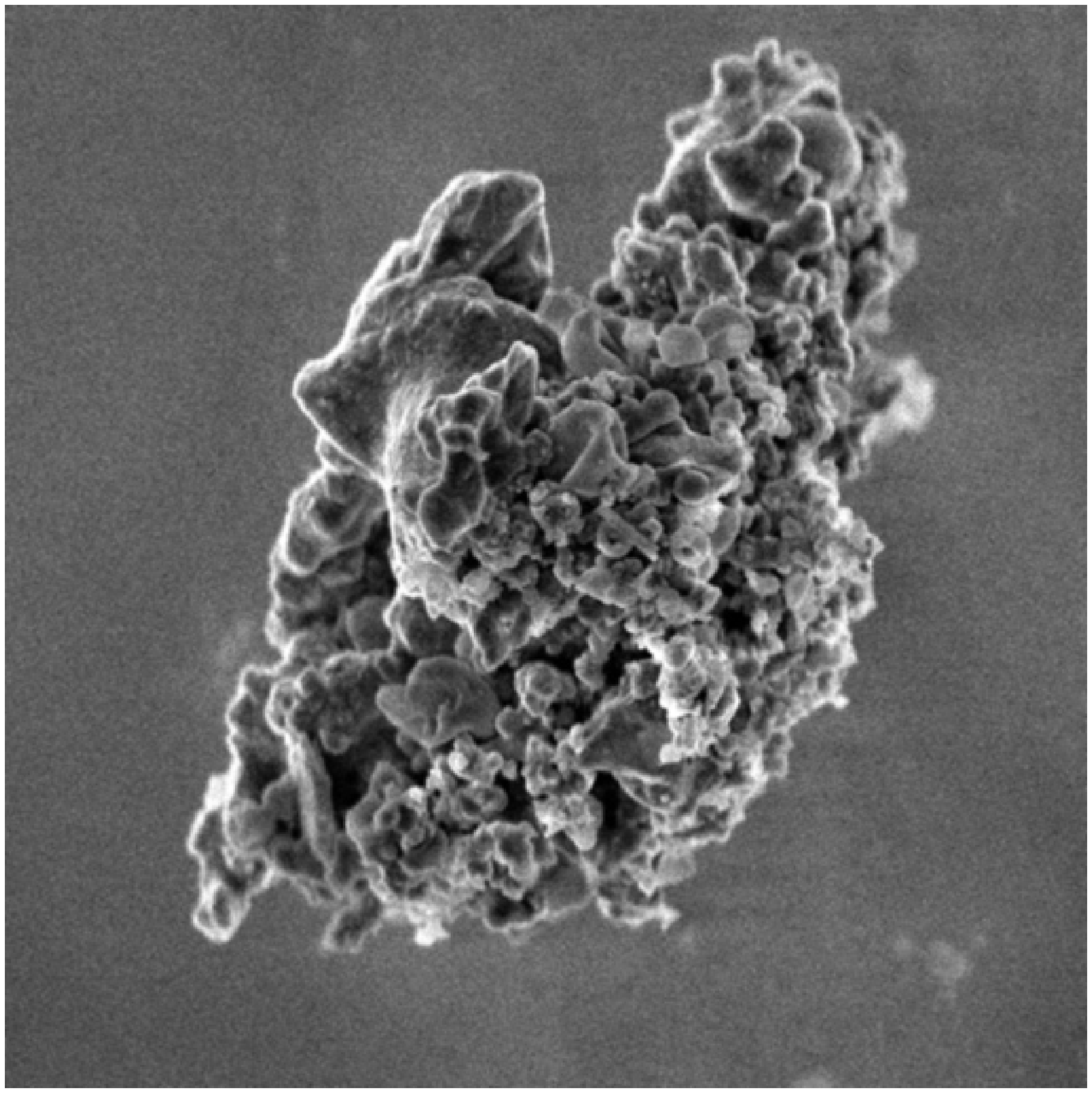}
\caption{Interplanetary dust grain found in the stratosphere,
which exhibits the typical porous structure, but at a much larger scale.
The sizes of interplanetary dust grains range from several $\mu$m
to 100 $\mu$m.  
Acknowledgement is made to NASA for allowing reproduction
of this picture from web site
http://stardust.jpl.nasa.gov.
}
\label{fig:1}
\end{figure}
\end{center}

The formation and evolution of interstellar dust grains are complex 
processes which involve
accretion and chemical reactions of impinging atoms and molecules, 
grain-grain collision and coagulation, 
photolysis and alteration by UV starlight, 
X-rays and cosmic rays,
erosion by sputtering 
and vaporization  
\citep{Draine2003}.
Due to the complexity of these processes,
there is no complete model
that accounts for all the relevant properties
of interstellar dust grains. 
The known properties are based on
a combination of 
astronomical observations, 
theoretical studies 
and analysis of meteorites and
solar-system dust.
Useful insight about the porosity of interstellar dust grains can
be obtained from its effect on
the observed extinction curves
\citep{Wright1987,Jones1988,Mathis1996,Snow1996,Fogel1998,Wolff1998,Schnaiter1999,Vadia1999,Krivova2000,Wurm2000,Iati2001}. 
Interplanetary dust particles and meteorites, 
collected in the upper atmosphere of Earth 
\citep{Brownlee1985},
were found to be fluffy and porous 
(Fig. \ref{fig:1}). 
Although these particles are larger 
than the typical interstellar dust grains, 
it is likely that
similar processes are 
responsible for their formation. 
Theoretical and experimental simulations of cosmic dust formation, 
through 
aggregation of molecular clusters,
produce fluffy and porous 
structures 
with amorphous surface morphology.

In most calculations of grain temperatures in PDRs, it was
assumed that grains are compact 
and can be considered as spherical.
However, several 
studies considered the
temperatures of non-spherical 
as well as porous grains
\citep{Greenberg1971,Blanco1980,Voshchinnikov1999,Voshchinnikov2005}.  
It was found that porous grains are significantly
colder, in most cases, than predicted by models which use compact 
spherical grains. 
While models of spherical grains predict grain temperatures in the 
range of $30 - 50$ K,
the actual temperatures of porous interstellar grains
may thus be significantly lower. 
In most favorable cases, the grain temperatures may go down to
the range of $15 - 30$ K for large grains and to $10 - 20$  
K for small grains 
\citep{Blanco1980}. 

It was
suggested long ago that 
micro-pores in interstellar dust grains
may trap gas particles and affect the
formation of molecules
\citep{Abadi1976}.
Recent experiment on 
ice surfaces at low temperatures ($10 - 25$ K) 
have shown that the porous morphology 
increases the desorption energy barriers
of adsorbed hydrogen atoms and molecules
\citep{Pirronello1999,Manico2001,Roser2002,Hornekaer2003,Hornekaer2005,Perets2005,Viti2004,Collings2004}. 
Analysis of some of these experiments shows that 
on such surfaces H$_2$ would form efficiently only at low 
surface temperatures of up to 20 K
\citep{Perets2005},
resembling the results obtained for olivine and carbon
\citep{Katz1999,Cazaux2004}. 
In PDRs, the relevant grain surfaces are expected to consist of 
bare silicate and carbon, not covered by ice mantles.
Below we examine the role of grain porosity in extending the
efficiency range of H$_2$ formation toward higher temperatures.

\section{H$_2$ formation on porous dust grains}
\label{sec:model}

\begin{center}
\begin{figure} 
\includegraphics[width=6in]{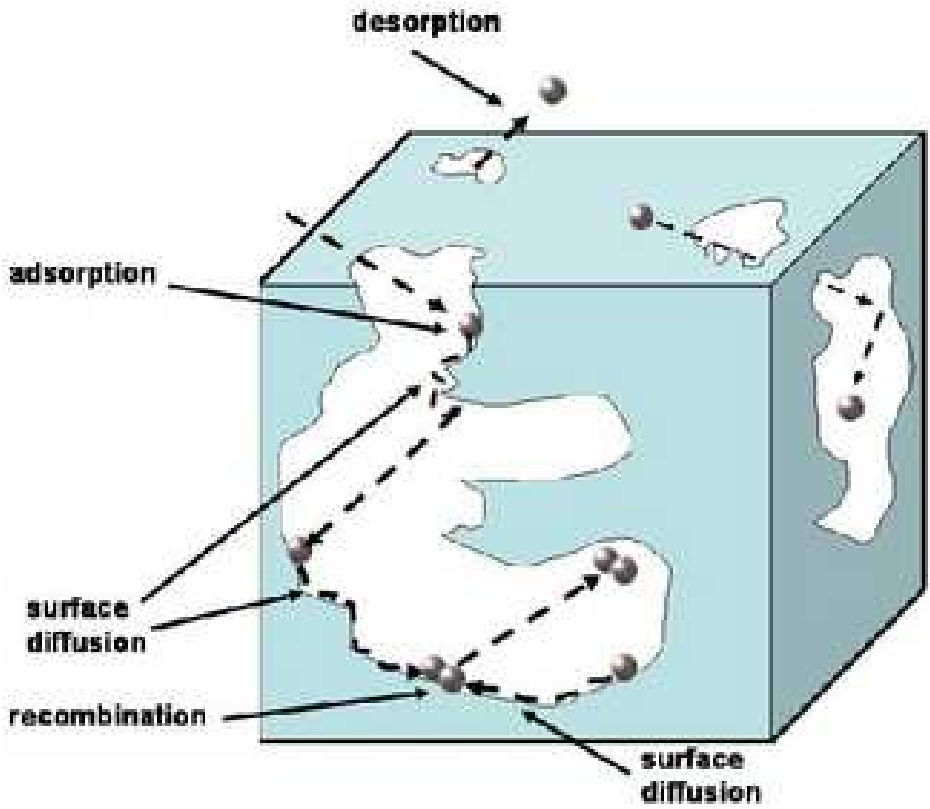}
\caption{
Schematic description of H diffusion and desorption 
in the internal and external surfaces of
a porous dust grain.
}
\label{fig:2}
\end{figure}
\end{center}

To evaluate the formation of H$_2$ on porous grains we
introduce a suitable rate equation model.
In this model we make a distinction between
the external surface of the grain and the 
internal surfaces within the pores.
Atoms adsorbed on the 
external surface behave as in the previous models.
They may hop between adjacent sites or desorb.
In addition, they may enter the pores and slits 
in the surface.
In the internal surfaces, 
similar diffusion and desorption processes apply.
However, an atom which desorb from an internal surface,
may re-adsorb inside the pore, rather than leave the grain.
Such an atom may leave the grain only after it reaches the external surface,
through a series of diffusion and desorption moves
(Fig. \ref{fig:2}).
Detailed modeling of these processes requires knowledge
on the exact geometry of the grains and their surface morphology.
However, the most important effects of grain porosity
can be analyzed by a simple rate equation model. 
The model, described below, assumes high connectivity of the pores, 
namely, that most of the internal surface of 
the grain is connected 
\citep{Kimmel2001}
and is treated as a single 
surface.    
The model thus consists of
two coupled rate equations 

\begin{subeqnarray}
\label{eq:N}
&\dot N_{\rm ext}& = 
F 
-   \left(\frac{S_{\rm edge}}{S_{\rm ext}}\right) a N_{\rm ext}
+   \left(\frac{S_{\rm edge}}{S_{\rm in}}\right) a N_{\rm in} 
          - 2 \left(\frac{a}{S_{\rm ext}}\right) {N_{\rm ext}}^2 
          - W N_{\rm ext} 
\slabel{eq:Next} \\
&\dot N_{\rm in}& = \ \ \ \ 
 \left(\frac{S_{\rm edge}}{S_{\rm ext}}\right) a N_{\rm ext} -  
 \left(\frac{S_{\rm edge}}{S_{\rm in}}\right) a N_{\rm in}   
               -  2 \left(\frac{a}{S_{\rm in}}\right) {N_{\rm in}}^2,  
\slabel{eq:Nin} 
\end{subeqnarray}

\noindent
where
$N_{\rm ext}$ 
and 
$N_{\rm in}$
are the populations of H atoms on the external and internal surfaces,
respectively.  
The first term in Eq.
(\ref{eq:Next}) 
describes the incoming flux 
of H atoms.
The second and third terms in Eq. 
(\ref{eq:Next}) 
[and the corresponding first and second terms in Eq. (\ref{eq:Nin})] 
describe the diffusion of atoms between the external surface and 
the pores, where $a$ is the hopping rate
given by 
Eq.
(\ref{eq:A}).
The numbers of adsorption sites on the
external and the internal surfaces of the grain are
$S_{\rm ext}$
and
$S_{\rm in}$,
respectively.
In addition, there is a small number, 
$S_{edge}$,
of sites, located at
the edges of the pores, thus
connecting the internal and the external surfaces of the grain.
An adsorbed atom moving between the internal and the external
surfaces must pass through one of these edge sites. 
The fourth term in 
Eq. (\ref{eq:Next}) 
corresponds to recombination of atoms 
on the external surface. 
Similarly, the third term in 
Eq. (\ref{eq:Nin}) 
corresponds to the recombination of atoms 
on the internal surface, inside the pores.
The last term in 
Eq. (\ref{eq:Next}) 
describes the desorption of atoms into the gas phase,
where
$W$ is the desorption rate given by
Eq.
(\ref{eq:W}).

To examine the effect of porosity on molecular hydrogen formation, 
we performed numerical integration of  
Eqs. (\ref{eq:N}) 
under steady state 
conditions
for grains of diameter
$d=10^{-5}$ (cm).
The porosity was characterized by the ratio between the 
internal and external surfaces,
$S_{in}/S_{ext}$.
The formation rate of molecular hydrogen per grain is given by 

\begin{equation}
R_{\rm grain} =   \left(\frac{a}{S_{\rm ext}}\right)  N_{\rm ext}^2 
          + \left(\frac{a}{S_{\rm in}}\right)   N_{\rm in}^2. 
\end{equation}

\begin{center}
\begin{figure} 
\includegraphics[width=6in]{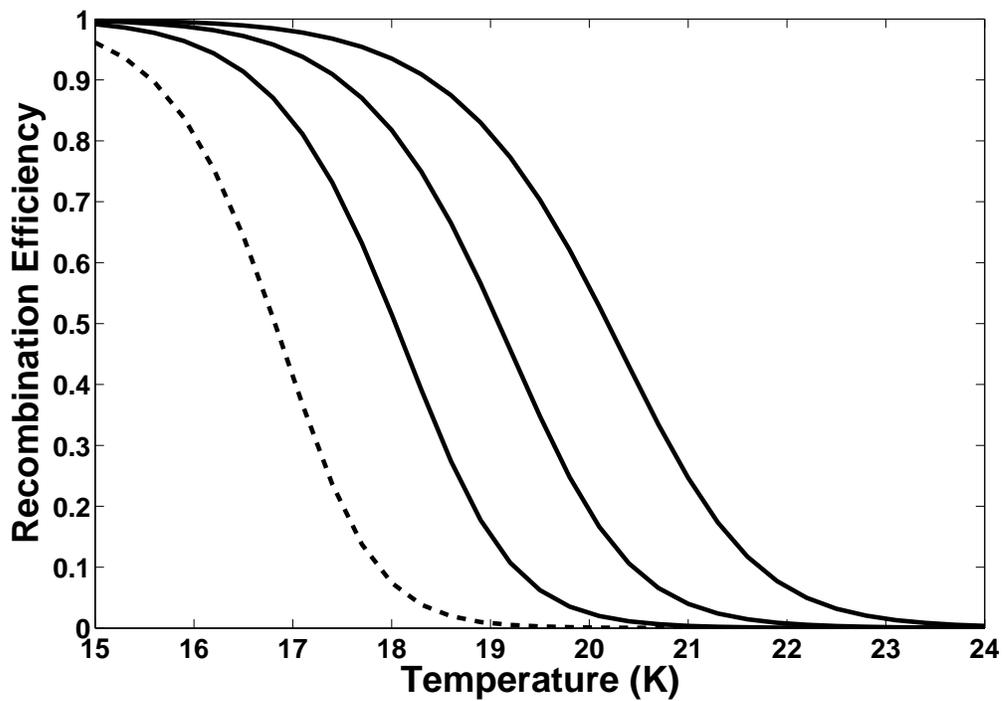}
\caption{H$_2$ recombination efficiancy 
for non-porous grains (dashed line) 
and for porous grains (solid lines) with 
$S_{in}/S_{ext}=$ 10, 100 and 1000. 
The more porous grains
exhibit higher bounds for efficient recombination.}
\label{fig:3}
\end{figure}
\end{center}

\noindent
In Fig. \ref{fig:3} we show the recombination efficiency of molecular hydrogen
on porous grains,
vs. grain temperature,
in the range between $10 - 35$ K.
Different levels of porosity are used,
namely
$S_{\rm in} / S_{\rm ext} = 10$, $100$ and $1000$. 
The results
show that porosity in interstellar grains
extends
the temperature range of high efficiency
toward higher temperatures. 
The extension of the temperature range is
logarithmic with the level of porosity
(Fig. \ref{fig:3}). 

The number of edge sites in the grain
is given by
$S_{\rm edge} / S_{\rm ext} = 5 \cdot 10^{-2}$.
We found that although
the edge sites
control the flow between the external and internal surfaces,
the recombination efficiency is only weakly dependent on
$S_{\rm edge}$.
In particular, we 
performed simulation with
$S_{\rm edge}$ 
values 10 times larger and smaller than
the value reported above. 
In both cases, the edges of the high-efficiency window were
only slightly shifted (not shown). 

\begin{center}
\begin{figure} 
\includegraphics[width=6in]{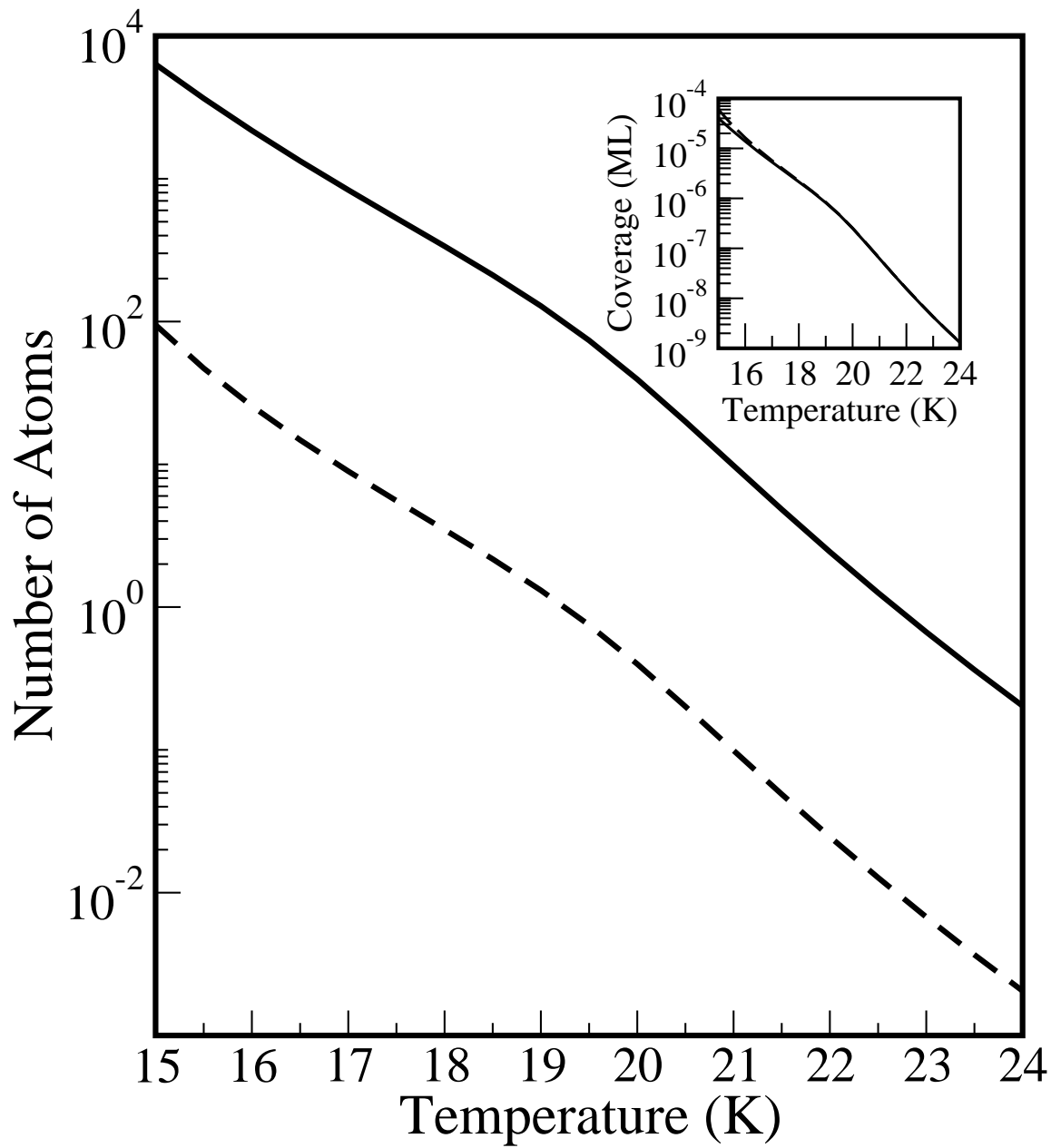}
\caption{The number of H atoms on the 
internal (solid line) and the external (dashed line)
surfaces of a
porous grain with
$S_{in}/S_{ext}=$ 100 as a function of the grain 
temperature. Most of the H$_2$ molecules form on 
the internal surface.
The inset shows the coverages of H atoms on the
internal (solid line) and the external (dashed line)
surfaces. These coverages turn out to be nearly 
identical.
}
\label{fig:4}
\end{figure}
\end{center}

In Fig. \ref{fig:4} 
we show the number of H atoms on the internal
and external surfaces (solid and dashed lines, respectively)
vs. temperature,
for a grain of diameter $d=10^{-5}$ (cm) where
$S_{\rm in}/S_{\rm ext}=100$.
It is found that for the relevant temperatures there is a large
population of H atoms in the pores, thus the rate equations
are suitable for the calculations presented here.
The inset in Fig. \ref{fig:4} 
shows the coverages of H atoms on the internal and on the
external surfaces. Both coverages are very small.
Since the desorption barrier of hydrogen molecules on the amorphous 
carbon surface is lower than the desorption barrier of atoms, the 
coverage of molecules is also expected to be small. 
This justifies the approximation made in this paper, 
ignoring the population of molecules on the surface and 
neglecting the LH rejection term.

\section{Discussion}
\label{sec:discussion}

The results presented above show that porosity extends the
range of efficient hydrogen recombination to higher temperatures.
This extension does not reach the grain temperatures of 30 - 50 K
expected in PDRs and thus cannot, by itself, explain the high
abundance of H$_2$ in these regions.
However, porous grains, particularly small ones, 
are expected to be colder and thus may be within the extended
temperature range of high efficiency.

The roughness of interstellar grain surfaces may broaden the
distribution of binding energies compared to the experimental
samples.
Thus, some of the adsorbed H atoms may be more strongly bound,
further extending the range of high efficiency towards higher
temperatures
\citep{Cuppen2005}.
Composite grains, which consist of both silicates and carbon
also exhibit a broader temperature window of high efficiency
\citep{Chang2005}.
Grains of diameters smaller than
$10^{-5}$ (cm) account for most of the surface area of grains
in interstellar clouds and may thus contribute significantly to
H$_2$ formation
\citep{Lipshtat2005}.
Temperature fluctuations may also reduce the average temperature
of small grains comapred to large ones. 

In addition to molecular hydrogen,
dozens of other molecular species have been observed in interstellar clouds.
Many of these molecules are formed through gas phase processes
not involving dust grains. 
Others are formed on dust-grain surfaces.
For example, certain organic molecules 
are believed to be formed on dust grains covered by ice mantles 
\citep{Tielens1997}.
The effect of dust morphology on H$_2$ formation may indicate
that it plays a role in other gas-grain processes.
Moreover, our results may 
have implications on the formation 
of molecules which have been previously thought to be formed mainly by gas
phase reactions. 
For example,
recent observations of interstellar N$_2$ have 
indicated much higher abundances
than predicted before 
\citep{Knauth2004}. 
It was suggested that these abundances may be explained 
by gas-grain reactions that had not been considered earlier. 
An analysis of these processes on porous grains may help 
to understand this new puzzle.

\section{Summary}
\label{sec:summary}           

We have considered the effect of porosity on the formation of
molecular hydrogen on interstellar dust grains.
The analysis was done using a rate equation model, which
takes into account both the internal and external surfaces 
of the grain.
It was found that porosity gives rise to
enhanced H$_2$ formation rates. 
It extends the range 
of high recombination efficiency toward higher temperatures.
This may contribute to the understanding of 
molecular hydrogen formation in PDRs,
which cannot be explained by models of compact dust grains.

\section{acknowledgments}

We thank E. Herbst, G. Vidali and V. Pirronello 
for helpful discussions.
This work was supported by the Israel Science Foundation
and the Adler Foundation for Space Research.

\clearpage
\newpage

\end{document}